\documentclass{elsart}
\usepackage{amssymb}
\usepackage{graphicx}
\usepackage{epsfig}
\def\tend{\mathop{\to}}
\def\lim{\mathop{\rm {lim}}}
\usepackage{natbib}
\begin{document}

\runauthor{Renat Kh. Gainutdinov, Aigul A. Mutygullina and Werner
Scheid}
\begin{frontmatter}
\title{Effects of nonlocality in time of interactions of an atom with
its surroundings on the broadening of spectral lines of atoms}
\author[Kazan]{Renat Kh. Gainutdinov}
\author[Kazan]{Aigul A. Mutygullina}
\author[Giessen]{Werner Scheid}

\address[Kazan]{Kazan State University, Department of Physics,
Kremlevskaya st.,18, Kazan, Russia}
\address[Giessen]{Institut f{\"u}r Theoretische Physik der
Justus-Liebig-Universit{\"a}t, D-35392, Giessen, Germany}

\begin{abstract}
We investigate effects of nonlocality in time of the interaction
of an atom with its surroundings on the spectral line broadening.
We show that these effects can be very significant: In some cases
nonlocality in time of this interaction can give rise to a
spectral line splitting.

PACS: 03.65.Bz; 32.70.Jz; 31.70.Hq.
\end{abstract}
\begin{keyword}
Nonlocal-in-time interaction, open quantum systems, spectral line
broadening
\end{keyword}

\end{frontmatter}

Separation of scales plays an important role in many problems of
physics because it allows one to select relevant degrees of
freedom and to describe a quantum system only in their terms.
Integrating out other degrees of freedom results in nonlocality in
time of the interaction in this system. As is well known, the
dynamics of such a system is not governed by the Schr{\"o}dinger
equation, since this equation is local in time, and interaction
Hamiltonians describe instantaneous interactions. At the same
time, in Ref. [1] it has been shown that the Schr{\"o}dinger
equation is not the most general dynamical equation consistent
with the current concepts of quantum physics, and a more general
equation of motion has been derived as a consequence of the
Feynman [2,3] and canonical approaches to quantum theory. Being
equivalent to the Schr{\"o}dinger equation in the case of
instantaneous interactions, this generalized dynamical equation
permits the generalization to the case where the dynamics of a
quantum system is generated by a nonlocal-in-time interaction. It
has been shown [1] that the generalized quantum dynamics (GQD)
developed in this way may be an important tool for solving various
problems in quantum theory.

An invaluable tool for computing physical quantities in the
theories with disparate energy scales are effective field theories
(EFT's) [4,5]. Following the pioneering work of Weinberg [6], the
EFT of nuclear forces has become very popular in nuclear physics
(for a rewiew, see [7]). To describe low energy processes
involving nucleons and pions, all operators consistent with the
symmetries of QCD are included in an effective Lagrangian. A
fundamental difficulty is that such Lagrangians yield graphs which
are divergent, and give rise to singular quantum mechanical
potentials. These potentials do not satisfy the requirements of
ordinary quantum mechanics and need to be regulated, and
renormalization must be performed. In this way one can
successfully perform calculations of many quantities in nuclear
physics. However, in such a way one cannot parametrize the $NN$
interaction by using some Hamiltonian, and there are not any
equations for renormalized amplitudes in subtractive EFT's. As a
result, in nuclear physics the Schr{\"o}dinger equation is not
valid even in the nonrelativistic limit. Meanwhile, as has been
shown in [1], only the generalized dynamical equation must be
valid in all the cases, not the Schr{\"o}dinger equation. In Ref.
[8] it has been shown that, in fact, low energy nucleon dynamics
in the EFT of nuclear forces is governed by the generalized
dynamical equation with nonlocal-in-time interaction operator when
this equation is not equivalent to the Schr{\"o}dinger equation.
Moreover, in leading order this dynamics is just the same as in
the case of the model [1,9] developed as a test model illustrating
the possibility of the extension of quantum dynamics provided by
the formalism of the GQD. This shows the predictive power of the
GQD: In Refs. [10,11] it has been shown that the existence of the
quark and gluon degrees of freedom should give rise to the fact
that low energy nucleon dynamics is governed by a nonlocal-in-time
interaction operator. In the case where the interaction is
separable the form of this operator is uniquely determined by the
corresponding form factor [1]. At the same time, as it follows
from the general analysis of the diagrams of the EFT, the leading
order two nucleon $T$ matrix is separable with the unite form
factor. Thus the GQD predicts, without summing diagrams and
performing procedures of regularization and renormalization, that
the leading order two nucleon $T$ matrix in the EFT approach
should be the same as in the above model with the unite form
factor, and this really takes a place. This predictive power of
the GQD is a result of the fact that the generalized dynamical
equation, which forms the basis of this formalism, has been
derived in Ref. [1] as the most general dynamical equation
consistent with the current concepts of quantum physics.

The above gives us the expect that the GQD can be applied  in the
theory of open quantum system where the separation of scales also
plays an important role. In this theory one consider an open
system in contact with its surroundings consist of a practically
infinite number of degrees of freedom and act as a whole identify,
referred as the thermal bath, on the open system. It is assumed
that the time scale of the open system is very long compared with
the relaxation time of the bath, but shorter than the recurrence
time for the whole system considered a closed system. For
describing the dynamics of open quantum systems one can start with
a closed system. A reduction to a small subsystem (open system)
produces nonlocal-in-time equations. The application of these
equations to real situations is extremely complex. For this reason
it is natural to describe the evolution of an open system only in
terms of its degrees of freedom by using a dynamical equation
derived from the first principles of quantum theory. Such a
program has been realized, for example, in the Lindblad theory
[12,13]. The possibility of the application of this theory to some
practical problems was considered, for example, in the works
[14-17]. However, the dynamical equation of this theory is a
Markovian-type equation, i.e., it does not take into account
nonlocality  of the interaction in time. Note, in this connection,
that the Markovian assumption is usually used to overcome the
above mentioned difficulties in the theory. For this reason
Lindblad theory is based on the axioms which imply that the theory
is local in time. This allows one to derive, as a consequence of
the general principles, an equation, which governs the evolution
of an open system, without entering into details of the physics of
the external system. The aim of the present paper is to show that,
keeping these advantages of Lindblad's theory, the GQD allows one
to describe the evolution of open quantum systems without using
the Markovian assumption. By using the example of a two-level atom
interacting with its environment, we will investigate the effects
of the nonlocality in time of the interaction on the character of
the dynamics of an open system.

Let us briefly consider the main features of the GQD. The basic
concept of the canonical formalism of quantum theory is that it
can be formulated in terms of vectors of a Hilbert space and
operators acting on this space. The postulates  establish the
connection between the vectors and operators and states of a
quantum system and observables, and prescribe how  the probability
of an event should be computed (see, for example [18]). In the
canonical formalism  these postulates are used together with the
dynamical postulate according to which the time evolution of a
quantum system is governed by the Schr{\"o}dinger equation. In the
equivalent  Feynman formalism quantum theory is formulated in
terms of probability amplitudes without resorting to vectors and
operators acting on a Hilbert space. In this approach the
following assumption is used as a basic postulate:

 The probability of an event is the absolute square of a
complex number called the probability amplitude. The joint
probability amplitude of a time-ordered sequence of events is the
product of separate probability amplitudes of each of these
events. The probability amplitude of an event which can happen in
several different ways is a sum of the probability amplitudes for
each of these ways.

 In particular, $\langle\psi_2| U(t,t_0)|\psi_1\rangle$, being the probability
 of finding,  the quantum system
 in the  state  $|\psi_2\rangle$ in a measurement at time $t$
 if at time $t_0$ it was in the state
 $|\psi_1\rangle$, can  be represented as a sum
of contributions from all alternative ways of realization of the
corresponding evolution process. Dividing these alternatives in
different classes, we can then analyze such a probability
amplitude in different ways [3]. For example, subprocesses with
definite instants of the beginning and  end of the interaction in
the system can be considered as such alternatives. In this way the
amplitude $\langle\psi_2|U(t,t_0)|\psi_1\rangle$  can be written
in the form [1]
\begin{equation}
\langle\psi_2| U(t,t_0)|\psi_1\rangle =
\langle\psi_2|\psi_1\rangle + \int_{t_0}^t dt_2 \int_{t_0}^{t_2}
dt_1 \langle\psi_2|\tilde S(t_2,t_1)|\psi_1\rangle,
\end{equation}
where $\langle\psi_2|\tilde S(t_2,t_1)|\psi_1\rangle dt_1dt_2$ is
the probability amplitude that at time $t_1$ the system in the
state $|\psi_1\rangle,$ then the interaction in the system  begins
in the time interval $(t_1,t_1+dt_1)$ and  ends in the time
interval $(t_2,t_2+dt_2)$  and at time $t_2+dt_2$ the system will
be in the state $|\psi_2\rangle.$ Eq. (1) is a consequence of the
first postulate of Feynman's approach to the quantum theory and,
therefore, derived from first principles of quantum theory.
According to the above assumption the probability amplitude
$\langle\psi_2|\tilde S(t_2,t_1)|\psi_1\rangle$ can itself be
represented by the sum  of amplitudes for each of the ways in
which the subprocess  with completely specified instants of the
beginning and end of the interaction in a quantum system can be
realized. However, some supplementary assumptions about the
history of the system are needed. In the Feynman approach it is
assumed that the history of the system can be represented by some
path in space-time. In this case the amplitude
$\langle\psi_2|\tilde S(t_2,t_1)|\psi_1\rangle$ can be in the form
of integrals over all paths corresponding to processes in which
the interaction begins at $t_1$ and ends at $t_2$. If, we also
assume that each path gives a contribution  is an exponential
whose (imaginary) phase is the classical action (in units of
$\hbar$) for this path (the second postulate of Feynman's theory)
and substitute the expression obtained in this manner into Eq.
(1), we arrive at Feynman's sum-over-paths formula for the
transition amplitudes. At the same time, in Ref. [1] it has been
shown that the use of the operator formalism of the canonical
approach allows one to derive a relation for the amplitudes
$\langle\psi_2|\tilde S(t_2,t_1)|\psi_1\rangle$, which can be used
as a dynamical equation without resorting to the supplementary
assumptions like the second postulate of Feynman's theory.

By using the operator formalism, we can represent the amplitudes
$\langle\psi_2|U(t_2,t_1)|\psi_1\rangle$ by the matrix elements of
the evolution operator $U(t,t_0)$, which must be unitary
\begin{equation}
U^{+}(t,t_0) U(t,t_0) = U(t,t_0) U^{+}(t,t_0) = {\bf 1}
\end{equation}
and satisfy the composition law
\begin{equation}
U(t,t') U(t',t_0) = U(t,t_0), \quad U(t_0,t_0) ={\bf 1}.
\end{equation}
At the same time, $\tilde S(t_2,t_1)$ whose matrix elements are
$\langle\psi_2|\tilde S(t_2,t_1)|\psi_1\rangle$ may be only an
operator-valued generalized function of $t_1$ and $t_2$ [1], since
only $U(t,t_0)={\bf 1}+ \int^{t}_{t_0} dt_2
\int^{t_2}_{t_0}dt_1\tilde S(t_2,t_1)$  must be an operator on the
Hilbert space. Nevertheless, it is convenient to call $\tilde
S(t_2,t_1)$ an "operator", using this word in generalized sense.
In the case of an isolated system the operator $\tilde S(t_2,t_1)$
can be represented in the form [1]
\begin{equation}
\tilde S(t_2,t_1) = exp(iH_0t_2) \tilde T(t_2-t_1) exp(-iH_0 t_1),
\end{equation}
where $H_0$ is the free Hamiltonian.

As has been shown in Ref. [1], for the evolution operator  given
by (1) to be unitary and satisfy the composition law (3), the
operator  $\tilde S(t_2,t_1)$ must satisfy the following equation
\begin{equation}
(t_2-t_1) \tilde S(t_2,t_1) = \int^{t_2}_{t_1} dt_4
\int^{t_4}_{t_1}dt_3 (t_4-t_3) \tilde S(t_2,t_4) \tilde
S(t_3,t_1). \label{main}
\end{equation}
With this equation one can obtain the operators $\tilde
S(t_2,t_1)$ for any $t_1$ and $t_2$, if the operators $\tilde
S(t'_2, t'_1)$ corresponding to infinitesimal duration times $\tau
= t'_2 -t'_1$ of interaction are known. It is natural to assume
that most of the contribution to the evolution operator in the
limit $t_2 \to t_1$ comes from the processes associated with the
fundamental interaction in the system under study. Denoting this
contribution by $H_{int}(t_2,t_1)$, we can write
\begin{equation}
\tilde{S}(t_2,t_1) \tend\limits_{t_2\rightarrow t_1}
H_{int}(t_2,t_1) + o(\tau^{\varepsilon}),\label{gran}
\end{equation}
where $\tau=t_2-t_1$. The parameter $\varepsilon$ is determined by
demanding that $H_{int}(t_2,t_1)$ must be so close to the solution
of Eq. (\ref{main}) in the limit $t_2\tend t_1$ that this equation
has a unique solution having the behavior (\ref{gran}) near the
point $t_2=t_1$.Thus this operator must satisfy the condition
\begin{equation}
 H_{int}(t_2,t_1)\tend\limits_{t_2 \tend t_1}
\int^{t_2}_{t_1} dt_4 \int^{t_4}_{t_1} dt_3
\frac{t_4-t_3}{t_2-t_1} H_{int}(t_2,t_4) H_{int}(t_3,t_1)+
o(\tau^{\epsilon}).\label{grh}
\end{equation}
Note that the value of the parameter $\epsilon$ depends on the
form of the operator $ H_{int}(t_2,t_1).$

Within the GQD the operator $H_{int}(t_2,t_1)$ plays the same role
as the interaction Hamiltonian  in the ordinary formulation of
quantum theory: It generates the dynamics of a system. Since this
operator is a generalization of the interaction Hamiltonian, we
call  this operator the generalized interaction operator. If
$H_{int}(t_2,t_1)$ is specified, Eq. (\ref{main}) allows one to
find the operator $\tilde S(t_2,t_1).$ Formula (1) can then be
used to construct the evolution operator $U(t,t_0)$ and
accordingly the state vector $|\psi(t)\rangle = |\psi(t_0)\rangle
+ \int_{t_0}^t dt_2 \int_{t_0}^{t_2} dt_1 \tilde S(t_2,t_1)
|\psi(t_0)\rangle $ at any time $t.$ Thus Eq. (\ref{main}) can be
regarded as an equation of motion for states of a quantum system.
It should be noted that Eq. (\ref{main}) is written only in terms
of the operators $\tilde S(t_2,t_1),$ and does not explicitly
contain operators describing the interaction in a quantum system.
It is a relation for $\tilde S(t_2,t_1)$ which contains the
contributions to the evolution operator from the processes with
specified instants of the beginning and end of the interaction in
the system. This relation is a unique consequence of the
composition law (3) and the representation (1) expressing the
Feynman superposition principle (the above assumption). For this
reason the relation (\ref{main}) must be satisfied in all the
cases. A remarkable feature of this fundamental relation is that
it works as a recurrence relation. For constructing the evolution
operator, it is sufficient to know the contributions to this
operator from the processes in which the duration time of
interaction is infinitesimal, i.e., from the processes ruled by
the fundamental interaction in the system. This makes it possible
to use the fundamental relation (\ref{main}) as a dynamical
equation. Its form does not depend on the specific features of the
interaction (the Schr{\"o}dinger equation, for example, contains
the interaction Hamiltonian). Since Eq. (\ref{main}) must be
satisfied in all the cases, it can be considered as the most
general dynamical equation consistent with the current concepts of
quantum theory. This generalized dynamical equation approaches the
Schr{\"o}dinger equation for appropriate boundary conditions.

By using (1), the evolution operator can be represented in the
form [1]
\begin{eqnarray}
\langle n_2|U(t,t_0)|n_1\rangle=  \langle n_2|n_1\rangle +
\frac{i}{2\pi}
\int^\infty_{-\infty} dx\exp[-i(z-E_{n_2})t]\nonumber\\
\times\frac { \langle n_2|T(z)|n_1\rangle \exp[i(z-E_{n_1})t_0]}
{(z-E_{n_2})(z-E_{n_1})}\label{evo}
\end{eqnarray}
 where $z=x+iy$,  and
$y>0$, $n$ stands for the entire set of discrete and continuous
variables that characterize the system in full, $|n\rangle$ are
the eigenvectors of the free Hamiltonian $H_0$, and $\langle
n_2|T(z)|n_1\rangle$ is defined by
\begin{equation}
\langle n_2|T(z)|n_1\rangle = i \int_{0}^{\infty} d\tau
\exp(iz\tau) \langle n_2|\tilde T(\tau)|n_1\rangle.\label{tz}
\end{equation}
The equation of motion (\ref{main}) is equivalent to the following
equation for the T matrix [1]:
\begin{equation}
\frac{d \langle n_2|T(z)|n_1\rangle}{dz} = -  \sum \limits_{n}
\frac{\langle n_2|T(z)|n\rangle\langle
n|T(z)|n_1\rangle}{(z-E_n)^2},\label{difer}
\end{equation}
with the boundary condition $$T(z) \tend \limits_{|z|\tend\infty}
i \int_{0}^{\infty} d\tau \exp(iz\tau) H_{int}^{(s)}(\tau),$$
where $H^{(s)}_{int}(t_2-t_1) = exp(-iH_0t_2) H_{int}(t_2,t_1)
exp(iH_0t_1)$ is the generalized interaction operator in the
Schr{\"o}dinger picture. As has been shown in Ref. [1], the
dynamics governed by Eq. (\ref{main}) is equivalent to the
Hamiltonian dynamics in the case where the generalized interaction
operator is of the form
\begin{equation}
 H_{int}(t_2,t_1) = - 2i \delta(t_2-t_1)
 H_{I}(t_1) .\label{loc}
\end{equation}
Here $H_{I}(t_1)$ is the interaction Hamiltonian in the
interaction picture. In this case the evolution operator given by
(1) satisfies the Schr{\"o}dinger equation. The delta-function
 in (\ref{loc}) emphasizes the fact that in this
case the fundamental interaction is instantaneous. At the same
time, Eq. (\ref{main}) permits the generalization to the case
where the interaction generating the dynamics of a system is
nonlocal in time [1]. In Ref. [1,9] this point was demonstrated on
exactly solvable models.

Let us consider an atom with its radiation field as an open
quantum system interacting with its surroundings. We will consider
the surroundings as a whole identity, referred as the thermal
bath.  Let the bath be non-Gaussian-Markovian, i.e., the  time
scale $t_c$ of the bath with respect to the system dynamics is
very short but $t_c\neq 0$. Further, we  assume that the evolution
operator $U_{op}(t_2,t_1)$ defined on the subspace ${\emph
H}_{op}$ describing the open system satisfies the semi-group law
\begin{equation}
U_{op}(t_2,t) U_{op}(t,t_1) = U_{op}(t_2,t_1), \quad t_2\geq t\geq
t_1,
\end{equation}
and $\Vert U_{op}(t_2,t_1)\Vert \leq 1$ for all $t_2>t_1$.
Obviously, due to a loss of probability from the open system, the
evolution operator is not unitary. Thus open quantum systems are
unstable. An approach to the theory of unstable  quantum systems
based on the semi-group law has been developed in the works of
Williams [19], Sinha [20] and Horwitz et all [21] (for reviews see
Ref. [22]). It is assumed that the evolution operator
$U_{op}(t_2,t_1)$ satisfies the semi-group law and is strongly
continuous. If this assumption can be seen as fulfilled from the
physical point of view, the above approach provides a fundamental
description of the dynamics of open quantum systems. In fact from
Stone's theorem [23] it follows that in the case of a closed
system the assumption that the evolution operator satisfies the
composition law (3), and is unitary and strongly continuous is
equivalent to the assumption that the evolution of a quantum
system is governed by the Schr{\"o}dinger equation [24]. This
result can be generalized to the case of open quantum systems.
However, the semi-group law (12) is not satisfied in general. It
is satisfied only in the case when at any time during the
evolution the system remains in a pure state belonging to the
subspace ${\emph H}_{op}$. But this, of course, does not take
place in the general case. This fact by itself allows us to
conclude that a semi-group law cannot hold for the evolution
operator on the Hilbert space ${\emph H}_{op}$, while the
evolution operator acting on the Banach space of density operators
satisfies the semi- group law in the general case. Thus the theory
has to be reformulated in the language of the density operator.
This program has been realized in the Lindblad theory. This theory
is founded on the same assumptions as the above approach
translated into the language of the density operator, i.e., on the
assumptions that the evolution operator on the Banach space of
density operators satisfies the semi-group law and is continuous.
However in this theory  the interaction generating the dynamics of
a quantum system is instantaneous. However, the interaction with
the non-Gaussian-Markovian bath must be nonlocal in time.

The GQD allows the extension of quantum theory to the case when
the evolution operator is not continuous, and precisely in this
case the interaction in a system is nonlocal in time [1]. As for
the validity of the semi-group law (11), the GQD deals with the
probability amplitudes in the spirit of the Feynman approach, and
the representation (\ref{main}) can be written for probability
amplitudes of any event concerning the evolution of the system. In
this case the operator $U(t_2,t_1)$ should be defined as an
operator whose matrix elements represent these amplitudes. In the
case under consideration we can define the matrix element $\langle
\psi_2|U_{op}(t_2,t_1)|\psi_1\rangle$ as the probability amplitude
that if at time $t_1$ the system was in the state
$|\psi_1\rangle\in{\emph H}_{op}$, then all the time between $t_1$
and $t_2$ the system will be in states belonging to the subspace
${\emph H}_{op}$, and at time $t_2$ the system will be found in
the state $|\psi_2\rangle\in {\emph H}_{op}$. The fact that in the
time interval $(t_2,t_1)$ the system is only in states belonging
${\emph H}_{op}$ means that all the time between $t_1$ and $t_2$
we deal with  the atom but not with its decay products (nuclear or
ions and electrons). Thus the matrix elements $\langle
\psi_2|U_{op}(t_2,t_1)|\psi_1\rangle$ represent some alternative
way of realization of the evolution process. The operator
$U_{op}(t_2,t_1)$ defined in this way is not equivalent to the
ordinary evolution operator defined on the space ${\emph H}_{op}$.
Since the operator $U_{op}(t_2,t_1)$ satisfies the semi-group law
(12), we can describe the system in terms of state vectors  on the
space ${\emph H}_{op}$. The operator $U_{op}(t,t_0)$  can then be
represented in the form
\begin{equation}
\langle \psi_2| U_{op}(t,t_0)|\psi_1\rangle = \langle
\psi_2|\psi_1\rangle + \int_{t_0}^t dt_2 \int_{t_0}^{t_2} dt_1
\langle \psi_2|\tilde S_{op}(t_2,t_1)|\psi_1\rangle,
\end{equation}
where the operator $\tilde S_{op}(t_2,t_1)$ describes the
contribution to the operator $U_{op}(t,t_0)$ from the process in
which the interaction in the system begins at time $t_1$ and ends
at time $t_2$. If the condition of the bath does not change in
time, the evolution of the open system under study must be
 invariant in time, and consequently the operator $\tilde
S_{op}(t_2,t_1)$ can be represented in the form $ \tilde
S_{op}(t_2,t_1) = exp(iH_0t_2) \tilde T(t_2-t_1) exp(-iH_0 t_1)$,
and the corresponding $T(z)$ is defined by (\ref{tz}). Here we use
the same notation for the operators $\tilde T(\tau)$ as in the
case of closed systems.

The operator $U_{op}(t_2,t_1)$ characterizes the evolution process
in the open quantum system. As example, let us consider  a
two-level atom. The matrix element $\langle {\bf k},{\bf
{\varepsilon}}_\lambda,1|U_{op }(t_2,t_1)|2\rangle$, is the
probability amplitude that if at time $t_1$ the system was in the
state $|2\rangle$, then at time $t_2$ the system will be found in
the state $ |{\bf k},{\bf {\varepsilon}}_\lambda,1\rangle$. Here
$|2\rangle$ and $|1\rangle$ denote the excited and  ground states
of the atom, respectively, ${\bf k}$ is the momentum of a photon
and ${\bf {\varepsilon}}_\lambda$ being the photon polarization.
If the ground state $|1\rangle$ of the atom can be regarded as
stable, then $\sum_{\lambda=1,2}\vert \langle {\bf k},{\bf
{\varepsilon}}_\lambda,1|U_{op}(\infty,0)|2\rangle\vert^2d\omega$
is the probability of a photon being emitted by a single atom with
energy in the interval $(\omega, \omega+d\omega)$.  From this it
follows that the spectral-line profile is described by the formula
\begin{equation}
\frac{dW_{21}(\omega)}{d\omega}=\sum_{\lambda=1,2}\vert \langle
{\bf k},{\bf
{\varepsilon}}_\lambda,1|U_{op}(\infty,0)|2\rangle\vert^2.
\end{equation}
Thus the operator  $U_{op}(t_2,t_1)$ determines the broadening of
spectral lines of atoms caused by the interaction with their
environment.  Let the generalized interaction operator describing
the interaction of the atom with its own radiation field and the
bath be of the form
\begin{equation}
H_{int}^{(s)}(\tau)=-2i\delta(\tau)H_I^{(v)}+H_{int}^{(b)}(\tau),
\end{equation}
with
$$H_I^{(v)}=\int d^3xj_\mu({\bf x},0)A^\mu({\bf x},0).$$
Here $j_\mu(x)$ is the current density operator, $A_\mu(x)$ is the
electromagnetic field potential and $H_{int}^{(b)}(t)$ is as
follows:
\begin{equation}
\left\{
\begin{array}{rcl}
\langle l|H^{(b)}_{int}(\tau)|j\rangle=\left (
f(\tau)-2i\Lambda\delta(\tau)\right ) \delta_{l2}\delta_{j2},\\
\langle {\bf
k},{\bf{\varepsilon}}_\lambda,l|H^{(b)}_{int}(\tau)|j\rangle=
\psi({\bf k})f(\tau)\delta_{l1}\delta_{j2},\\
\langle l|H^{(b)}_{int}(\tau)|{\bf k},{\bf
{\varepsilon}}_\lambda,j\rangle=\psi^*({\bf k})
f(\tau)\delta_{l2}\delta_{j1}, \\
\langle {\bf k}_2,{\bf
\varepsilon}_{\lambda_2},l|H^{(b)}_{int}(\tau)|{\bf k}_1,{\bf
\varepsilon}_{\lambda_1},j\rangle=\psi({\bf k}_2)\psi^*({\bf
k}_1)f_1(\tau)\delta_{l1}\delta_{j1},\\
\end{array}
\right.
\end{equation}
where the form factor $\psi({\bf k})$ is chosen as $\psi({\bf
k})=c_1(d^2+k^2)^{-\frac{1}{2}}$, and $\Lambda$, $c_1$ and $d$
being some constants. The function $f(\tau)$, characterizing the
nonlocality in time of the interaction,  is not arbitrary since
the interaction operator (15) must satisfy the condition
(\ref{grh}). For  this condition to be satisfied, the function
$f(\tau)$ must be of the form
\begin{equation}
f(\tau)=-\frac{i}{2\pi}\int_{-\infty}^\infty
exp(-iz\tau)\left(\frac{b_1}{\ln(-z)}+\frac
{b_2}{\ln^2(-z)}\right)dz,
\end{equation}
where $b_1=-\frac{1}{4\pi c_1^2}$ and only one parameter $b_2$ is
free.  The operator $H_I^{(v)}$ describes the interaction of an
isolated atom with the electromagnetic field, while the operator
$H_{int}^{(b)}(\tau)$ describes the interaction in which the bath
manifests itself.

The main assumption that we have used in choosing the form of the
operator $H_{int}^{(b)}(\tau)$ is that its matrix element $\langle
2|H^{(b)}_{int}(\tau)|{\bf k},{\bf \varepsilon}_\lambda,1\rangle$
should be of the form $\langle 2|H^{(b)}_{int}(\tau)|{\bf k},{\bf
\varepsilon}_\lambda,1\rangle=\psi^*({\bf k})f(\tau)$. As has been
shown in [1], there is a one-to-one correspondence between the
form of the interaction operator and the ultraviolet (UV) behavior
of the matrix elements of the evolution operator as functions of
momenta of photons: If this behavior satisfies the requirements of
conventional quantum theory, then $H_{int}(t_2,t_1)$ must
necessarily be of the form (\ref{loc}), i.e., the interaction must
be instantaneous, and, on the contrary, if this behavior is "bad"
and leads to UV divergences, then $H_{int}(t_2,t_1)$ must be
nonlocal in time. In the case of separable interactions the above
behavior and hence the form $H_{int}(t_2,t_1)$ are determined by
the UV behavior of the corresponding form factors. Let the form
factor has the following UV behavior $\psi({\bf
k})\sim\frac{1}{|{\bf k}|^\alpha}$. In the case $\alpha>1$, the
function $f(\tau)$ must be of the form $f(\tau)=A\delta(\tau)$,
where $A$ being some constant, i.e., the interaction is
instantaneous. In the case $0<\alpha\leq 1$ (the restriction
$\alpha>0$ is necessary for the generalized dynamical equation
(\ref{main}) to have the mathematical meaning) the function
$f(\tau)$ has no such singularity as the delta-function at
$\tau=0$, and hence the interaction is nonlocal in time. Since the
interaction of the atom with its surroundings is nonlocal in time,
 we have to use a form factor corresponding to the case
$0<\alpha\leq 1$. The interaction is nonlocal in time for any
$0<\alpha\leq 1$. However the case $\alpha=1$ seems to be most
natural. In fact, in this case the UV behavior of the form factors
gives rise to the logarithmic divergences in the Born series,
i.e., to the one of the fundamental divergences of QED. The
simplest form factor with such a UV behavior is of the form
$\psi({\bf k})=c_1(d^2+k^2)^{-\frac{1}{2}}$ (the parameter $d$ is
needed in order to resolve the problem of the infrared
divergences). For these reasons we have chosen the form factor in
this form. From the condition (\ref{grh}) it follows that the
operator $H_{int}^{(s)}(\tau)$ must also describe the processes of
the atom-photon scattering, and the corresponding matrix element
must be of the form $\langle {\bf k}_2, {\bf
\varepsilon_{\lambda_2}},1|H_{int}^{(s)}(\tau)|{\bf k}_1, {\bf
\varepsilon_{\lambda_1}},1\rangle=\psi({\bf k}_2)\psi^*({\bf
k}_1)f(\tau)$ with the same duration-time function $f(\tau)$ given
by (17). Note that there is only one free parameter $b_2$ in this
function. Thus the form of the interaction operator is determined
by the UV behavior of the form factor $\psi({\bf k})$. However,
this concerns only the nonlocal part of the interaction operator,
and one may supplement the interaction operator by an
instantaneous part [9]. The interaction operator
$H_{int}^{(b)}(\tau)$ also contains the instantaneous term. But in
our model it describes only the interaction of the atom in the
excited state. We do not take into account this interaction of the
atom in the ground state, because we assume that the possibility
for a photon decay is very small when the atom is in the ground
state, and hence the interaction can lead only to an energy shift
of this state.

Let us consider the case where $c_1=0$, i.e., where the
interaction of the atom with its environment is reduced to the
instantaneous interaction being described by the potential
$\langle i|V|j\rangle =\Lambda\delta_{i2}\delta_{j2}$. We will
assume that $\langle i|V|j\rangle $ is such that the interaction
of the atom with the radiation field can be considered as a small
perturbation, and hence the problem can be solved by expanding in
powers of $H_{I}^{(v)}$. In the first order, for $\langle {\bf
k},{\bf \varepsilon}_\lambda,1|T(z)|2\rangle $, we have
\begin{equation}
\langle {\bf k},{\bf \varepsilon}_\lambda,1|T^{(1)}(z)|2\rangle
=\frac{\langle {\bf k},{\bf
\varepsilon}_\lambda,1|H_{I}^{(v)}|2\rangle
(z-E_2)}{z-E_2-\Lambda}.
\end{equation}
Substituting this expression into (8) and using (14), we then get
\begin{equation}
\frac{dW_{21}(\omega)}{d\omega}=A\omega\sum\limits_{\lambda=1,2}\int
d\Omega_k\left |\frac{\langle {\bf k},{\bf
\varepsilon}_\lambda,1|H_{I}^{(v)}|2\rangle
}{E_1+\omega-E_2-\Delta E_2+\frac{i}{2}\Gamma_2}\right |,
\end{equation}
where $\Delta E_2=Re\Lambda,$ $\Gamma_2=-2 Im\Lambda$, and $A$ is
a renormalization factor. In the model under consideration
$\langle i|V|j\rangle $ can be considered as the corresponding
matrix elements of an potential which parametrize effects of
impacts of the atom with particles from its surroundings, and
hence $\Delta E_2$ and $\Gamma_j$ can be interpreted, accordingly,
as the impact line shift and width.

Thus in the case where the interaction with its surroundings is
instantaneous, our model yields the same results as the theory of
 impact broadening of spectral lines. At the same time, our model
allows the extension to the case where the interaction of the atom
with its surroundings results not only in radiationless
transitions between the excited and ground states, but also in the
emission of photons. Such processes are generally nonlocal in
time, and the parameter $c_1$ cannot be zero. Let us now solve the
problem in this case. We will assume that the interaction of the
atom with its surroundings is such that we can restrict ourselves
to the zero order in powers of $H_{I}^{(v)}$. In this case  the
dynamical equation (\ref{difer}) can be exactly solved, and the
corresponding solution is
\begin{equation}
\left\{ \begin{array}{rcl}
\langle l|T(z)|j\rangle=t_{22}(z)\delta_{l2}\delta_{j1},\\
 \langle {\bf k},{\bf
\varepsilon}_\lambda,2|T(z)|1\rangle=\psi({\bf k})t_{12}(z),\\
\langle 2|T(z)|{\bf k},{\bf
\varepsilon}_\lambda,1\rangle=\psi^*({\bf k})
t_{21}(z), \\
\langle {\bf k}_2,{\bf \varepsilon}_{\lambda_2},1|T(z)|{\bf
k}_1,{\bf \varepsilon}_{\lambda_1},1\rangle=\psi({\bf
k}_2)\psi^*({\bf
k}_1)t_{11}(z),\\
\end{array}
\right.
\end{equation}
where
\begin{equation}
t_{22}(z)=\frac{\left(\Lambda+T_1(z)\right)(z-E_2)}{z-E_2-\Lambda-T_1(z)},
\quad
t_{11}(z)=\frac{T_1(z)(z-E_2-\Lambda)}{z-E_2-\Lambda-T_1(z)},
\end{equation}
\begin{equation}
t_{21}(z)=t_{12}(z)=\frac{T_1(z)(z-E_2)}{z-E_2-\Lambda-T_1(z)},
\end{equation}
with
$$
T_1(z)=\frac{-\frac{1}{4\pi c_1^2}\frac{d^2+(z-E_1)^2}{(z-E_1)^2}}
{-\pi i+\ln(\frac{z-E_1}{d_0})+\frac{d^2\ln
\frac{d}{d_0}}{(z-E_1)^2}-\frac{\pi d}{2(z-E_1)}
-\frac{b_2(d^2+(z-E_1)^2)}{b_1(z-E_1)^2}},
$$
where $d_0=1 {\textsf{eV}}$. By using this formula and Eq.
(\ref{evo}) we can construct the operator $U_{op}(t,t_0)$. For
example, we can obtain the matrix element $\langle {\bf k},{\bf
{\varepsilon}}_\lambda,1|U(\infty,0)|2\rangle$, which, as it
follows from (14), determines the broadening of the spectral line
of the atom. By using (14) and the above expression for the T
matrix, we then get
\begin{equation}
\frac{dW_{21}(\omega)}{d\omega}=\frac{1}{2\pi}\frac {A f(\omega)}
{\left(\omega+E_1-E_2-\Delta E(\omega)\right)^2+\frac{1}{4}
\Gamma^2(\omega)},
\end{equation}
 where
$$
f(\omega)=\frac{d^2+\omega^2}{4\pi c_1^2\omega^4}
\left(\pi^2+\left[\ln
\frac{\omega}{d_0}+\frac{d^2}{\omega^2}\ln\frac{d}{d_0}-\frac{\pi
d}{2\omega}
-\frac{b_2(d^2+\omega^2)}{b_1\omega^2}\right]^2\right)^{-1};
$$
$$
\Delta E(\omega)=Re\Lambda-f(\omega)\left(\omega^2\ln
\frac{\omega}{d_0}+d^2 \ln \frac{d}{d_0}
-\frac{\pi}{2}d\omega-\frac{b_2}{b_1}\left[d^2+\omega^2\right]\right);
$$
$$
\Gamma(\omega)=2\left(Im\Lambda+\pi\omega^2 f(\omega)\right).
$$

This formula describes the spectral line profile of the atom in
our model. In order to find out how the nonlocality in time of
interaction of the atom with its surroundings can influence on the
spectral line broadening, we have made the calculations of the
profile (23) for various values of the parameters of the model.
The results of the calculations are given in Fig.1.

For certain values of the parameters the nonlocality of
interaction has very significant effects on the
 spectral line broadening. Fig.1(c,d), for example, shows
that in some cases such a nonlocality of the interaction gives
rise to the splitting of the spectral line of the atom.

\begin{figure}
\begin{center}
 \begin{tabular}{cc}
   \psfig{figure=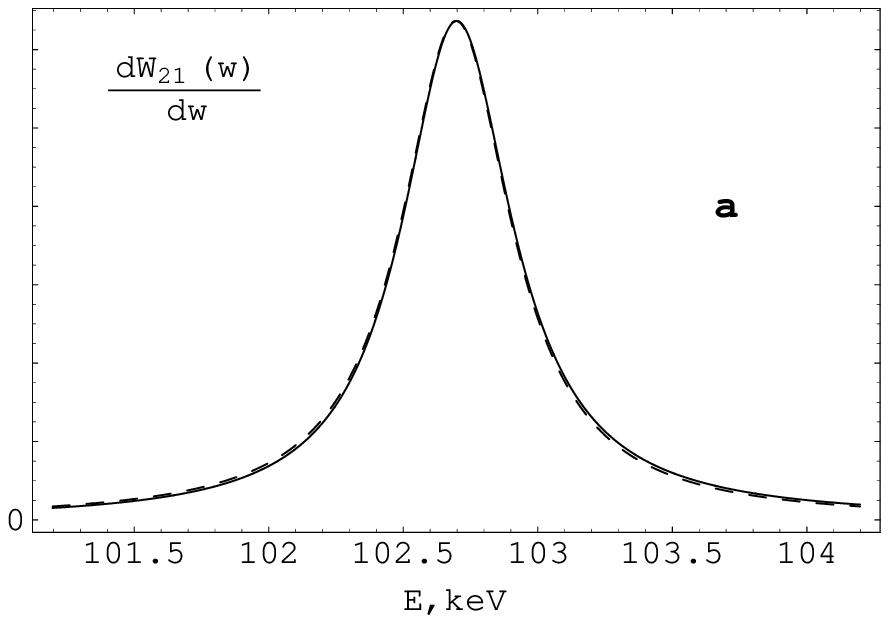,height=5cm}  &
\psfig{figure=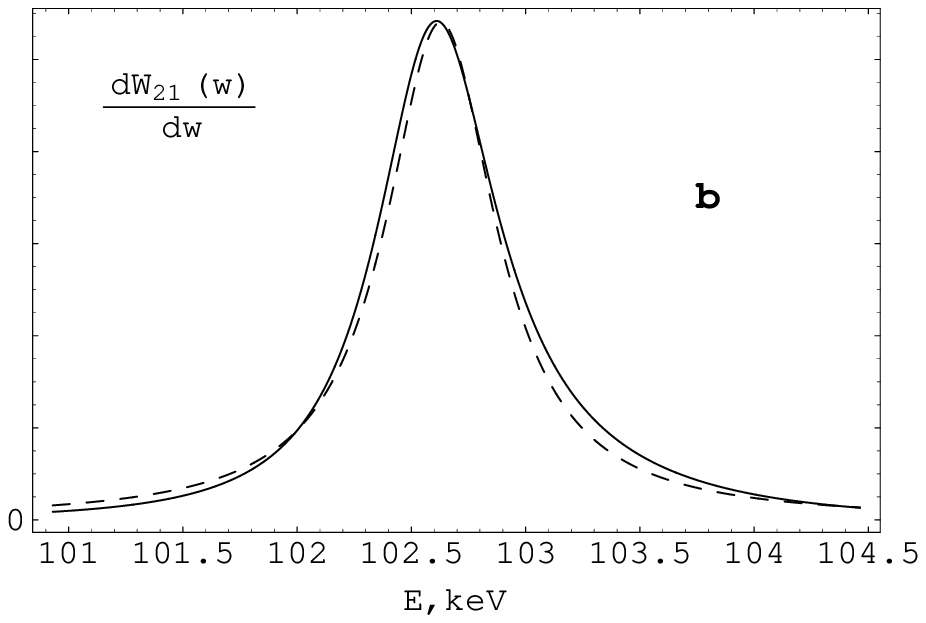,height=5cm}\\
\psfig{figure=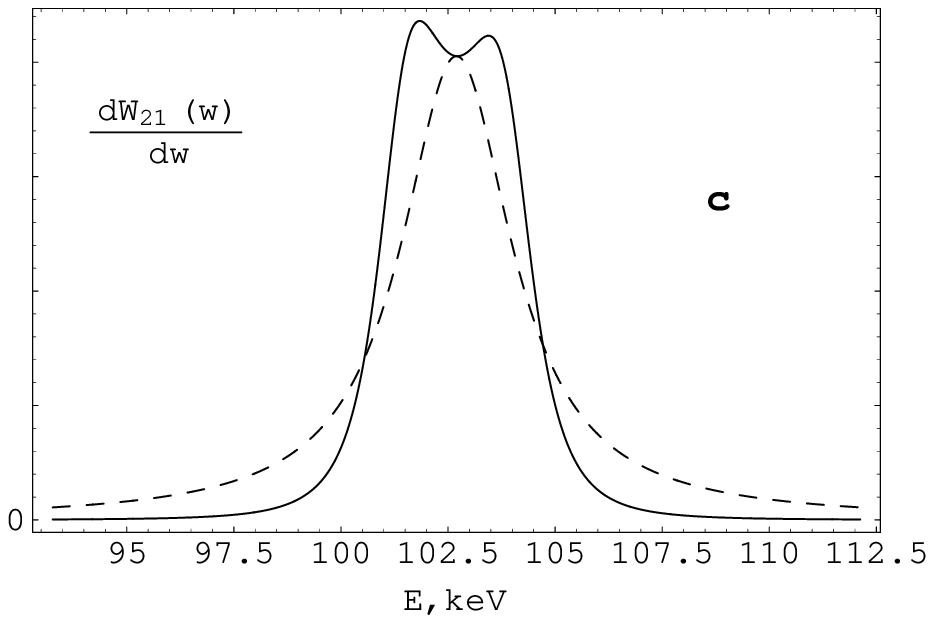,height=5cm}  &
\psfig{figure=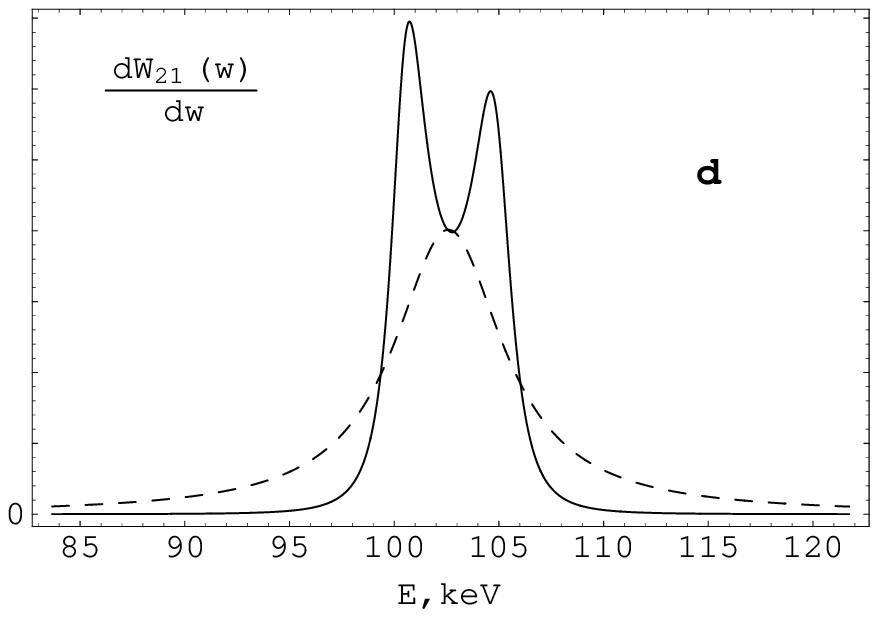,height=5cm}
  \end{tabular}
 \end{center}
\caption{The profile of the spectral line $2P_{\frac{1}{2}}\to
1S_{\frac{1}{2}}$ of the hydrogen-like atom, $Z=92$,
$E_2-E_1=102697$ $eV$. The dashed line represent the corresponding
Lorentzian line for the parameters given in Table 1.}
     \end{figure}

      \begin{table}
\caption{The parameters of the interaction operator (15)}
\begin{center}
\begin{tabular}{|l| c |c |c| c|}
\hline Figs., & $\Lambda$, $eV$ & $c_1^2$, $eV^{-1}$&$d$, $eV$&
$b_2, eV$\\
\hline
a & 250$\imath$  & 25 & $1.25\cdot 10^7 $& -0.052\\
b & 250$\imath$  & 9 & $1.25 \cdot 10^7 $& -0.144\\
c & 300$\imath$  & 0.25 &$ 1.15 \cdot 10^7$ & -5.170\\
d & 250$\imath$  & 0.09 & $1.05 \cdot 10^7$ & -14.281 \\
\hline
\end{tabular}
\end{center}
\end{table}
To summarize and to conclude, we hade shown that the GQD can be
used to describe, in a natural way, the evolution of an open
quantum system whose interaction with its environment is nonlocal
in time. We have defined the operator $U_{op}(t_2,t_1)$, in terms
of which the dynamics of open quantum systems can be described in
the same way as the dynamics of closed systems. In the case when
the open system is an atom with its radiation field this operator
determines the broadening of spectral-line profiles of atoms
caused by a nonlocal-in-time interaction with its environment. By
using an exactly solvable model, we found that the nonlocality in
time of the interaction of the atom with its surroundings can have
substantial effects on the spectral line broadening. As it follows
from our calculations, for the most values of the parameters the
broadening of the spectral line profile does not differ from the
Lorentzian line, i.e., the effects of nonlocality in time of
interaction on the broadening of the spectral line are negligible.
However, there is a narrow interval of values of the parameters
for which the broadening of the spectral line differs profoundly
from the Lorentzian, and the spectral lines can even split. This
means that some resonance condition of the bath can exists  which
has the effect that the spectral lines split. Our results can be
considered as a prediction of the above effect.

\end{document}